\documentclass[superscriptaddress, aps, showpacs, floatfix, prb,twocolumn]{revtex4-2}

\usepackage{graphicx}
\usepackage{dcolumn}
\usepackage{bm}

\newcommand{\MSS}{MnSc$_2$S$_4$\,}

\begin{document}

\title{Spin excitations in the magnetically ordered phases of \MSS}

\author{B. T{\'o}th}
\email{toth.boglarka@ttk.bme.hu}
\affiliation{Department of Physics, Institute of Physics, Budapest University of Technology and Economics, M\H{u}egyetem rkp. 3. H-1111, Budapest, Hungary}

\author{K. Amelin}
\author{T. R{\~o}{\~o}m}
\author{U. Nagel}
\affiliation{National Institute of Chemical Physics and Biophysics, Akadeemia tee 23, 12618 Tallinn, Estonia}

\author{V. Tsurkan}
\author{L. Prodan}
\affiliation{Experimental Physics V, Center for Electronic Correlations and Magnetism, Institute of Physics, University of Augsburg, 86159 Augsburg, Germany}
\affiliation{Institute of Applied Physics, 5 Academiei str. MD-2028 Chisinau, Republic of Moldova}

\author{I. K\'ezsm\'arki}
\affiliation{Experimental Physics V, Center for Electronic Correlations and Magnetism, Institute of Physics, University of Augsburg, 86159 Augsburg, Germany}

\author{S. Bord{\'a}cs}
\affiliation{Department of Physics, Institute of Physics, Budapest University of Technology and Economics, M\H{u}egyetem rkp. 3. H-1111, Budapest, Hungary}

\begin{abstract}
Recent neutron scattering experiments suggested that frustrated magnetic interactions give rise to antiferromagnetic spiral and fractional skyrmion lattice phases in \MSS. Here, to trace the signatures of these modulated phases, we studied the spin excitations of \MSS by THz spectroscopy at 300\,mK up to 12\,T.  We found a single magnetic resonance with linearly increasing frequency in field. The corresponding $g$-factor of Mn$^{2+}$ ions $g$\,=\,1.96, and the absence of other resonances imply very weak anisotropies and negligible contribution of higher harmonics to the spiral state. The significant difference between the dc magnetic susceptibility and the lowest-frequency ac susceptibility in our experiment implies the existence of mode(s) below 100\,GHz. 
\end{abstract}

\maketitle

\section{Introduction}
There has been a continued interest in materials with competing magnetic interactions as they may give rise to highly correlated fluctuating states \cite{Yamashita2008, Shimizu2003} or exotic magnetic orders \cite{Lee2000, Bramwell2001, Hiroaki2005, Intro_to_frustr2011}. To minimize the magnetic energy, as a compromise, spin spirals described by a single $\mathbf{q}$-vector often emerge in these compounds with frustrated interactions \cite{Luttinger1946, Izyumov1984, Kimura2003, Yamasaki2006, Kimura2006, Seki2008}. Recent theoretical works on frustrated magnets indicated that quartic-order terms in the Landau free energy can even stabilize a rich variety of multi-$\mathbf{q}$ spin states, including topologically non-trivial magnetic skyrmions \cite{Kawamura2012,Leonov2015,Hayami2022}. The helicity and vorticity of skyrmions can fluctuate, allowing the manipulation of these internal degrees of freedom, which may couple to electric polarization \cite{Leonov2015,Hayami2022}. Very recently, nanoscale skyrmions were detected in centrosymmetric Gd magnets with triangular \cite{Kurumaji2019}, kagome \cite{Hirschberger2019} and square lattices \cite{Khanh2020}, initializing experimental studies of magnetic skyrmions in frustrated magnets. 

The bipartite diamond lattice can also become frustrated with competing nearest ($J_1$) and next-nearest neighbor ($J_2$) interactions \cite{Bergman2007,Chen2009}. \MSS realizes this exchange-frustrated model, as the magnetic Mn$^{2+}$ ions with $S$=\,5/2 spins occupy the diamond sublattice of its spinel structure \cite{Fritsch2004}. The Curie-Weiss temperature $\Theta_{CW}=-$22.9\,K is an order of magnitude larger than the magnetic ordering temperature $T_N$=\,2.3\,K, implying that the magnetic interactions are strongly frustrated in this compound \cite{Fritsch2004,Gao2017}. According to neutron scattering experiments, above the ordering temperature $T_N$, the fluctuations are correlated and a unique spiral spin liquid state emerges \cite{Gao2017}. This state orders into a sinusoidally modulated collinear phase at $T_N$, which becomes incommensurate below 1.64\,K and, finally, transforms to a helical state below 1.46\,K. Furthermore, elastic and inelastic neutron scattering experiments suggested that in a finite magnetic-field, a triple-$\mathbf{q}$ state is stabilized, which, based on Monte Carlo simulations, is associated with a fractional antiferromagnetic skyrmion lattice \cite{Gao2020}.

The spectroscopy of the magnetic resonances has been proven to provide valuable information on the magnetic order and allows accurate determination of microscopic interaction parameters \cite{Spinwave2018}. Since the spin spiral has a periodicity larger than that of the chemical unit cell, there is a folding of the spin-wave dispersion into the smaller Brillouin zone of the spin spiral, a series of excitations may emerge in the $\Gamma$-point (k=0), that can be probed by absorption spectroscopy with high energy resolutions \cite{Kataoka1987,Katsura2007,deSousa2008,Fishman2019}. Such modes corresponding to the distortion of the phase or collective tilt of the plane of the spiral have been found, for example, in orthorhombic manganites \cite{Takahashi2012}, BiFeO$_3$ \cite{Cazayous2008,Talbayev2011,Nagel2013}, cubic chiral helimagnets \cite{Date1977,Schwarze2015}, and in Cu$_2$OSeO$_3$ \cite{Onose2012}. The higher dimensional multi-$\mathbf{q}$ states may give rise to additional modes as in the case of skyrmion lattice, where a breathing,  a clockwise and a counterclockwise rotational modes were predicted and observed \cite{Mochizuki2012,Onose2012,Schwarze2015}.

In this paper, we study the magnetic-field dependence of the spin excitations in \MSS by THz spectroscopy up to 17\,T. We carried out the experiments in the paramagnetic phase at 2.5\,K, as well as in the ordered state at 300\,mK. At 300\,mK the zero-field ground state is the helical spiral state and in the 4.5\,--\,7\,T field range the tripe-$\mathbf{q}$ state is expected to emerge \cite{Gao2017}. We detected a single resonance at both temperatures, whose frequency is linearly proportional to the magnetic field, characterized by a $g$-factor $g$=1.96. Within the measured frequency range of the experiments, we did not observe further excitations in either case. These findings indicate weak magnetic anisotropy of the Mn spins and also weak coupling of additional modes of the modulated magnetic structures to oscillating magnetic and electric fields of the THz radiation.

\section{Methods}
Single crystals with a typical size of $\sim$1\,mm$^3$ were grown by chemical transport technique, as described in Ref.~\onlinecite{Gao2017}. Several co-oriented crystals facing to the [111] direction were glued to obtain a mosaic with $\sim$2\,mm diameter and 0.65\,mm thickness. Subkelvin temperatures were reached in a modified Oxford TLE200 wet dilution refrigerator at the National Institute of Chemical Physics and Biophysics (KBFI), Tallinn. The propagation vector of the exciting unpolarized light was parallel to the external magnetic field, which is the so-called Faraday configuration. 2.5\,K measurements were also performed at KBFI, on the TeslaFIR cryostat setup. These measurements were performed with polarized light, in Voigt configuration, i.e.~propagation vector of the exciting polarized light was perpendicular to the external magnetic field. In both cases, the spectra were measured with an SPS200 far-infrared Martin-Puplett interferometer with a 300\,mK silicon bolometer. 

The field-induced change in the absorption coefficient $\alpha$ was calculated as
\begin{equation}
    \alpha(B) - \alpha(0\,\mathrm{T})= -\frac{1}{d}\mathrm{ln}\left( \frac{\mathcal{I}(B)}{\mathcal{I}(0\,\mathrm{T})}\right),
\end{equation}
where $\mathcal{I}(B)$ is transmitted light intensity at a specific magnetic field $B$, and $d$ is the sample thickness.

\section{Results}

We measured light absorption of a \MSS mosaic in the far-infrared range, between 100\,GHz and 3\,THz at 2.5\,K, and between 100\,GHz and 2.1\,THz at 300\,mK.

\begin{figure}[h]
\centering
\includegraphics[width = \linewidth]{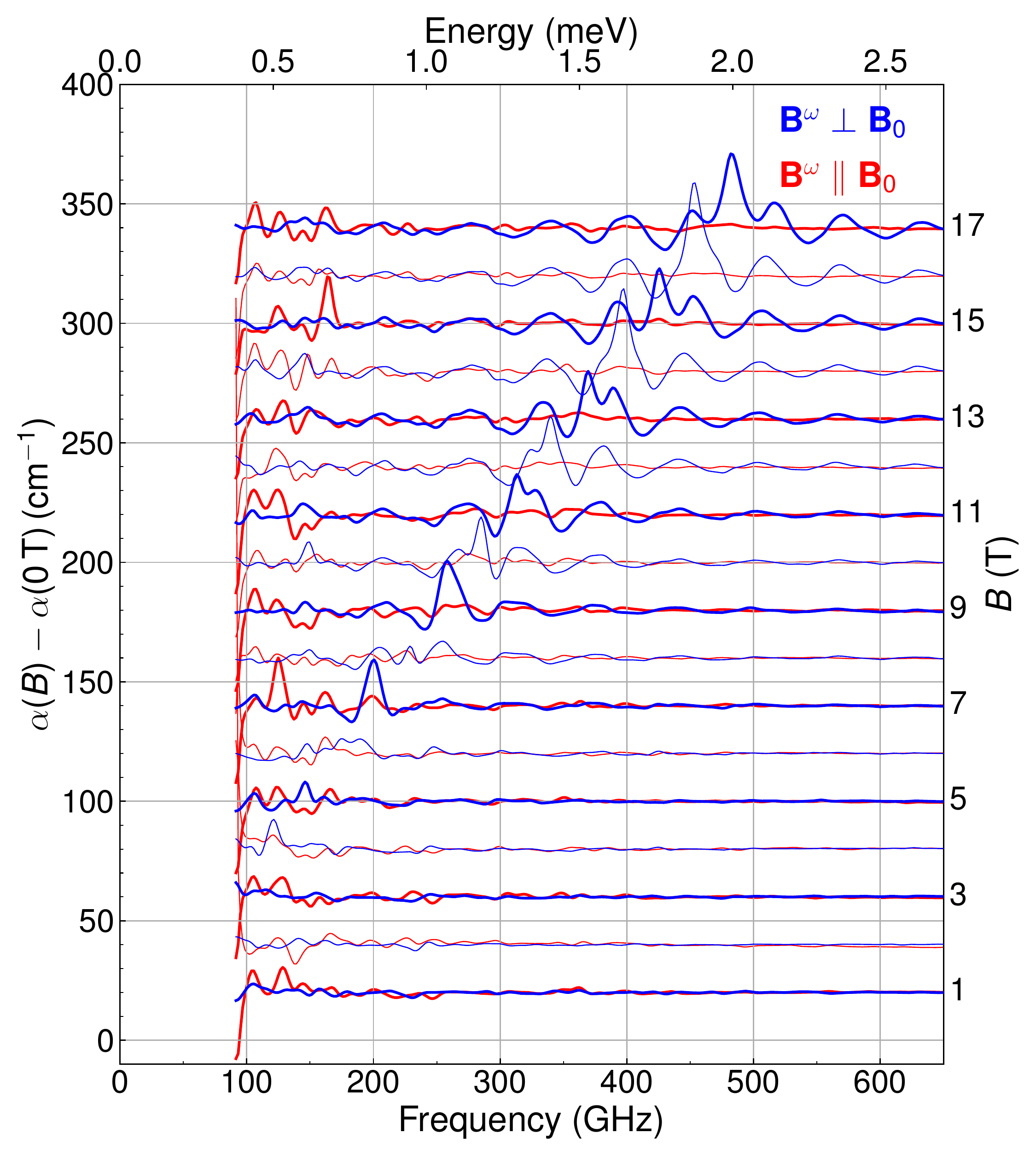}
\caption{Magnetic field and polarization dependence of the THz absorption spectrum measured in Voigt configuration in \MSS at $T=$\,2.5\,K. The absorption differences are shown with respect to zero field spectrum in magnetic fields up to 17\,T. Bold spectra were measured at odd field values. Undulation of the spectra in the vicinity of the resonance is due to multiple reflections within the plane-parallel sample that is distorted by the change of sample optical constants near the spin resonance mode.}
\label{fig:twopol}
\end{figure}

Figure~\ref{fig:twopol} shows the field dependence of the differential absorption spectra at 2.5\,K. We resolved a single paramagnetic resonance, which shifts linearly with field. The wavy baseline in the vicinity of the peak is caused by the magnetic-field-induced change of the interference pattern, arising due to multiple reflections in the nearly plane-parallel sample. A linear fit on the magnetic field dependence of the resonance frequency results in a 27\,$\pm$\,0.6\,GHz/T slope, and a zero-field offset of 20.4\,$\pm$\,7.2\,GHz. The slope corresponds to a $g$-factor of 1.93\,$\pm$\,0.05. The fact that we did not detect any deviation from the linear field dependence of the resonance apart from a small off-set suggests that there is a small anisotropy of spin Hamiltonian parameters. The small anisotropy is also consistent with the small deviation of $g$-factor from the free electron value, both caused by the spin-orbit coupling. The resonance line appears only when the direction of the alternating magnetic field is perpendicular to the external magnetic field, $\mathbf{B}^{\omega}\perp \mathbf{B}_{0}$, and is absent when $\mathbf{B}^{\omega}\parallel\mathbf{B}_{0}$, which is consistent with a simple paramagnetic behaviour.

\begin{figure}[h]
\centering
\includegraphics[width = \linewidth]{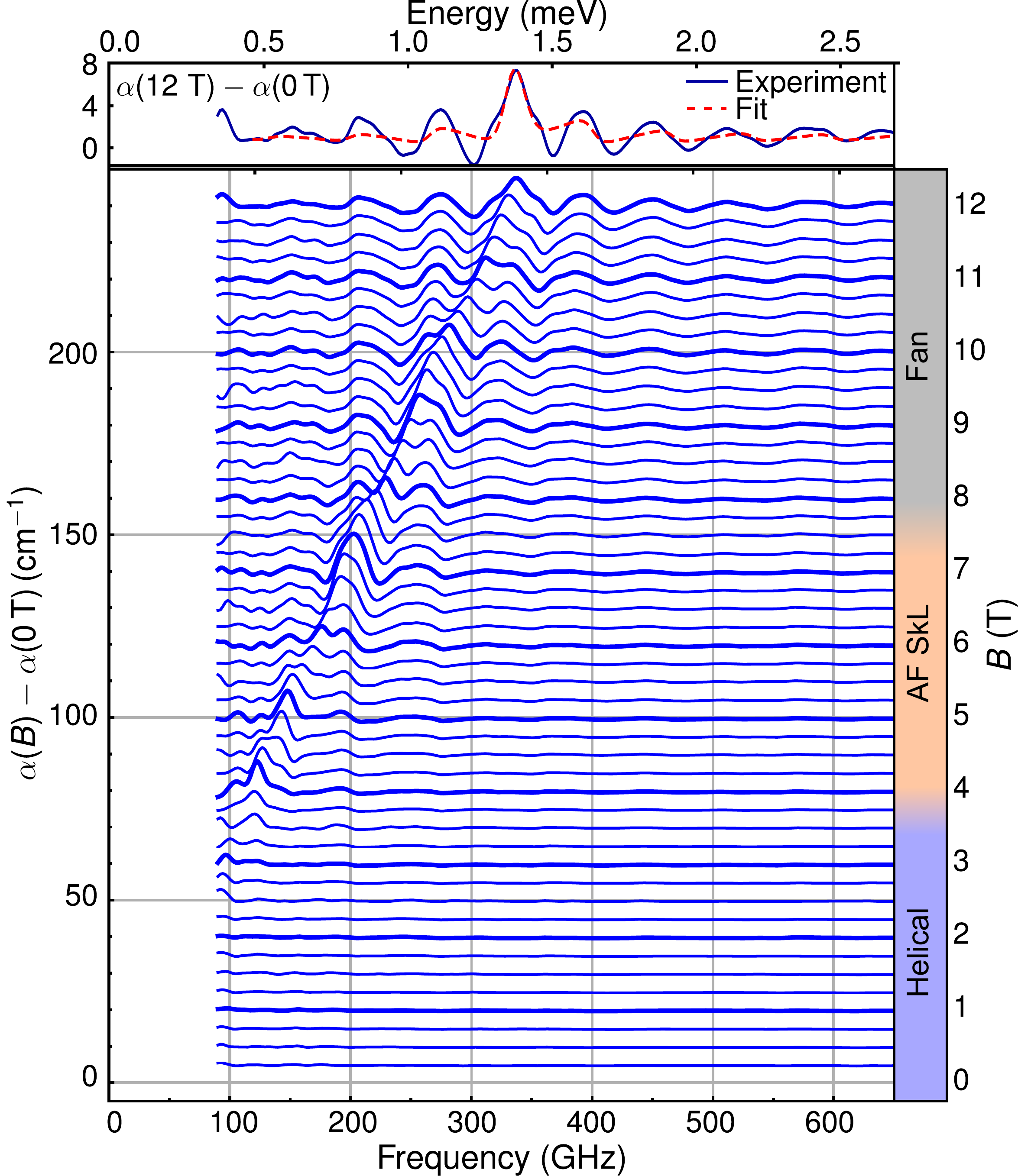}
\caption{THz absorption spectra in Faraday configuration for unpolarized radiation at $T$\,=\,300\,mK. In the bottom panel, the magnetic field dependence of the THz absorption spectrum is shown. Spectra measured in integer fields are plotted with thick lines. Top panel shows the spectrum measured in 12\,T (blue line) and its fit to Lorentzian model (red dashed line). Side bar shows the field regions where the magnetic phase has helical, antiferromagnetic skyrmion lattice (AF SkL), and fan structure according to Ref.\,\onlinecite{Gao2020}.}
\label{fig:waterfall}
\end{figure}

Figure.~\ref{fig:waterfall} shows the field dependence of the absorption spectra relative to zero field at 300\,mK. The spectra follow a similar magnetic-field dependence as the ones measured at 2.5\,K. The linear fit to the resonance peak positions shows a resonance shift of 27.5\,$\pm$\,0.4\,GHz/T, and a zero-field offset of 9.5\,$\pm$\,1.5\,GHz. The calculated slope corresponds to a $g$-factor of 1.96\,$\pm$\,0.02, which is the same as that of deduced in the paramagnetic phase within the error of the measurement. The finite frequency intercept is somewhat smaller as compared to the paramagnetic resonance at 2.5\,K.

Neither in the ordered nor in the paramagnetic phase could we resolve clear deviation from the linear field dependence. If there is any anisotropy induced gap, it is below 100\,GHz, the low-frequency cut-off of our experiment. The spin resonance detected at 300\,mK is not sensitive to the magnetic phase transitions that are suggested to occur at 4.5\,T and 7.5\,T, according to Ref.~\onlinecite{Gao2020}. Moreover, in our frequency window, we did not detect any other resonances, which may arise due to the emergence of a modulated spin structure, such as a spin spiral or magnetic skyrmion lattice. The absence of any signature of the modulated phase might correspond to the weak spin-orbit interaction and the related weak magnetic anisotropy of Mn$^{2+}$. Without a sizable magnetic anisotropy, the spin spiral is harmonic and the modes folded to the reduced Brillouin zone remain silent. The oscillation of the plane of the harmonic spiral may induce modulation of the electric polarization via the inverse Dzyaloshinskii-Moriya coupling \cite{Katsura2007,deSousa2008}, however, this mechanism is active only for spin cycloids, i.e., it cannot generate infrared active modes in the helical state of \MSS. These are the most likely reasons for not observing additional spin resonances in the covered spectral range. 

From the absorption spectrum, the $\omega$\,$\to$\,0 magnetic susceptibility $\chi$ can be obtained using the Kramers-Kronig relations, assuming that $\chi$ is small and the dielectric function $\varepsilon$ is constant in the THz range:
\begin{equation}
\chi(\omega \to 0) = \frac{2}{\pi}\frac{c}{\sqrt{\varepsilon}}\int_{0}^{\infty} \frac{\alpha (\omega)}{\omega^{2}} \mathrm{d}\omega,
\label{eq:KK}
\end{equation}

where $\alpha$ is the absorption coefficient, $\omega$ is the angular frequency, and $c$ is the speed of light in vacuum. We fitted the experimental $\alpha(B)-\alpha(0\,\mathrm{T})$ spectrum at $B$\,=\,12\,T with a single resonance. The magnetic susceptibility is described by a Lorentzian oscillator:
\begin{equation}
\chi(\omega) = \frac{S}{\omega_0^2-\omega^2-i\omega\gamma}
\end{equation}
where $\omega_0$ is the resonance frequency, $\gamma$ is the damping parameter and $S$ is the oscillator strength. To take into account multiple reflections within the sample, we modeled it as a Fabry-Perot etalon with infinite number of reflections. By assuming that the resonance is absent in zero field, our model provided an estimate for the THz dielectric constant: $\varepsilon$\,=\,12.1. The evaluation of the integral in Eq.~\ref{eq:KK} with the Lorentzian model gave $\chi(\omega \to 0)$\,=\,2.5$\times$10$^{-3}$ for fields oscillating perpendicular to the static field. This transverse susceptibility is an order of magnitude smaller than $\chi_0$\,=\,0.021, the value published for the longitudinal, static susceptibility in Ref.~\onlinecite{Fritsch2004}. Since the magnetization curve is nearly linear even in the magnetically ordered phases \cite{Gao2020}, and the anisotropy is weak, these transverse and longitudinal susceptibilities should be nearly equal in the static limit. The missing spectral weight, i.e., the difference between $\chi(\omega \to 0)$ and $\chi_0$, must lie outside of the frequency range of our measurement system, implying the presence of further resonance(s) below 100\,GHz, the low-frequency cutoff of the present study. In fact, an antiferromagnetic spiral emerging due to exchange frustration has three Goldstone modes in the absence of anisotropy: a phason mode, $\Phi_0$ corresponding to rotations within the plane of the spiral and two others, $\Psi_{\pm1}$ associated to out-of-plane rotations \cite{Fishman2019}. Magnetic anisotropy terms compatible with cubic symmetry may gap these modes making them detectable with microwave spectroscopy.

\section{Conclusions}

Earlier elastic and inelastic neutron scattering studies combined with Monte Carlo simulations found multiple phases in \MSS, including a multi-$\mathbf{q}$ state, such as an antiferromagnetic skyrmion phase. Motivated by these findings, we studied the magnetic field dependence of the spin excitations in \MSS by THz spectroscopy in the paramagnetic phase as well as in the ordered state. Although the material has a rich phase diagram with multiple modulated magnetic phases, we only observed a single resonance, whose frequency does not exhibit anomalies at the critical fields separating these phases. This resonance has $g$-factor close to 2 and shows no deviation from the linear field dependence, which indicates a small anisotropy. Other collective modes of the modulated states were not detected likely due to their negligible magnetic dipole activity, being the consequence of weak magnetic anisotropy. The analysis of the intensity suggests that further spin excitation(s) at lower frequencies should be present.

\begin{acknowledgments}
This work was supported by the Hungarian Academy of Sciences via Bolyai 00318/20/11 and by the Ministry of Innovation and Technology and the National Research, Development and Innovation Office within the Quantum Information National Laboratory of Hungary. Support was also given by the Estonian Ministry of Education and Research personal research funding PRG736 and  European Regional Development Fund Project No. TK134. The authors acknowledge the support of the bilateral program of the Estonian and Hungarian Academies of Sciences under the contract NKM 2018-47 and NKM 2021-24. This work was also partially supported by the Deutsche Forschungsgemeinschaft (DFG) through Transregional Research Collaboration TRR 80 (Augsburg, Munich, and Stuttgart) and by the project ANCD 20.80009.5007.19 (Moldova).
\end{acknowledgments}

\end{document}